\documentclass[english,a4paper,pre,twocolumn,amsmath,amssymb,superscriptaddress]{revtex4}

\usepackage{bmpsize}
\usepackage{graphicx,color}	
\usepackage[dvipsnames]{xcolor}						
\usepackage{amssymb}
\usepackage{natbib}
\usepackage{array}
\usepackage{subfigure}

\begin{document}

\title{Composition inversion in mixtures of binary colloids and polymer}

\author{Isla Zhang}
\affiliation{School of Chemistry, Cantock's Close, University of Bristol, BS8 1TS, UK}
\affiliation{Centre for Nanoscience and Quantum Information, Bristol BS8 1FD, UK}

\author{Rattachai Pinchaipat}
\affiliation{Centre for Nanoscience and Quantum Information, Bristol BS8 1FD, UK}
\affiliation{HH Wills Physics Laboratory, University of Bristol, Bristol BS8 1TL, UK}

\author{Nigel B. Wilding}
\affiliation{Department of Physics, University of Bath, Bath BA2 7AY, UK}

\author{Malcolm A. Faers}
\affiliation{Bayer AG, Alfred Nobel Str. 50, 40789, Monheim, Germany.}

\author{Paul Bartlett}
\affiliation{School of Chemistry, Cantock's Close, University of Bristol, BS8 1TS, UK}

\author{Robert Evans}
\affiliation{HH Wills Physics Laboratory, University of Bristol, Bristol BS8 1TL, UK}

\author{C. Patrick Royall}
\affiliation{School of Chemistry, Cantock's Close, University of Bristol, BS8 1TS, UK}
\affiliation{Centre for Nanoscience and Quantum Information, Bristol BS8 1FD, UK}
\affiliation{HH Wills Physics Laboratory, University of Bristol, Bristol BS8 1TL, UK}

\begin{abstract}
Understanding the phase behaviour of mixtures continues to pose challenges, even for systems that might be considered ``simple''. Here we consider a very simple mixture of two colloidal and one non-adsorbing polymer species which can be simplified even further to a size-asymmetrical binary mixture, in which the effective colloid-colloid interactions depend on the polymer concentration. We show that this basic system exhibits surprisingly rich phase behaviour. In particular, we enquire whether such a system features only a liquid-vapor phase separation (as in one-component colloid-polymer mixtures) or whether, additionally, liquid-liquid demixing of 
two colloidal phases can occur. Particle-resolved experiments show demixing-like behaviour, but when combined with bespoke Monte Carlo simulations, this proves illusory, and we reveal that only a single liquid-vapor transition occurs. Progressive migration of the small particles to the liquid phase as the polymer concentration increases gives rise to \emph{composition inversion} --- a maximum in the large particle concentration in the liquid phase. Near criticality the density fluctuations are found to be dominated by the larger colloids.
\end{abstract}

%\pacs{61.20.-p; 64.70.Dv; 61.20.Gy}

\maketitle

\section{Introduction}

It is difficult to overstate the importance of mixtures, as they constitute the vast majority of materials. The most basic mixtures are those of two species, and studies of such binary atomic and molecular mixtures have a distinguished history. In particular, much is known about the topologies of their possible phase diagrams from a theoretical perspective~\cite{rowlinson,vankonyenburg1976}. However, experimental studies of such systems do not generally provide detailed information on central features such as the compositions and structure of coexisting phases, the character of near-critical fluctuations, and the link between the form of the microscopic interactions and the phase behavior. In this respect colloidal dispersions are versatile systems for gaining insight into these basic aspects of phase behaviour \cite{frenkel2002,ivlev}.

Adding non-adsorbing polymer depletant induces entropically driven attraction between colloidal particles, and such mixtures can be interpreted as colloidal systems with the polymer degrees of freedom \emph{integrated out}. Because the resulting colloid-colloid interactions are very similar to those of atoms and molecules, these systems likewise exhibit fluid, liquid and crystalline phases, along with metastable states such as glasses and gels \cite{poon2002,lekkerkerker}. A key advantage over atomic and molecular systems is that one can readily modify the strength and range of the effective particle interactions, while directly observing the structure at the particle-resolved level using microscopy \cite{ivlev}.

Although colloid-polymer mixtures have proven invaluable in elucidating the properties of single component fluids \cite{poon2002,aarts2004,royall2007c}, surprisingly little attention has been given to binary colloidal mixtures. Investigations to date include sedimentation profiles where kinetics and equilibrium phase behaviour can exhibit a complex interplay \cite{leocmach2010,zhou2011pccp,zhou2011} and studies of dynamics where electrostatic interactions lead to non-intuitive behavior \cite{pandey2013}. Studies of phase behaviour indicate a time-dependent stratification of the sediment into layers with differing composition \cite{zhou2011pccp,zhou2011}. However, experience with single component colloids shows that complex behavior can occur under a gravitational field \cite{schmidt2004prl,delasheras2012}, and it is uncertain whether the stratification observed in \cite{zhou2011pccp,zhou2011} is thermodynamic or gravitational in origin.

The current theoretical understanding of such mixtures remains relatively little developed. One important result is that the free-volume approach of Lekkerkerker \emph{et al.} \cite{lekkerkerker1992} can be generalised to the present ternary mixture of two colloid species and polymer and predicts only vapor-liquid coexistence for the parameters of our experiments ~\cite{evans2013}. Here we report a combined experimental and simulation study of the phase behavior of a binary colloid mixture with a single species of added polymer. We describe the system in terms of an effective binary colloidal mixture in which the explicit polymer degrees of freedom are integrated out: colloidal interactions then take a form that is parameterized in terms of a polymer reservoir mass fraction $c_p^r$ which plays the role of inverse temperature. Central questions concern (\emph{i}) the topology of the phase diagram of such a system and (\emph{ii}) the structure and composition of the coexisting phases.

\begin{figure*}
\centering
\includegraphics[width=120mm]{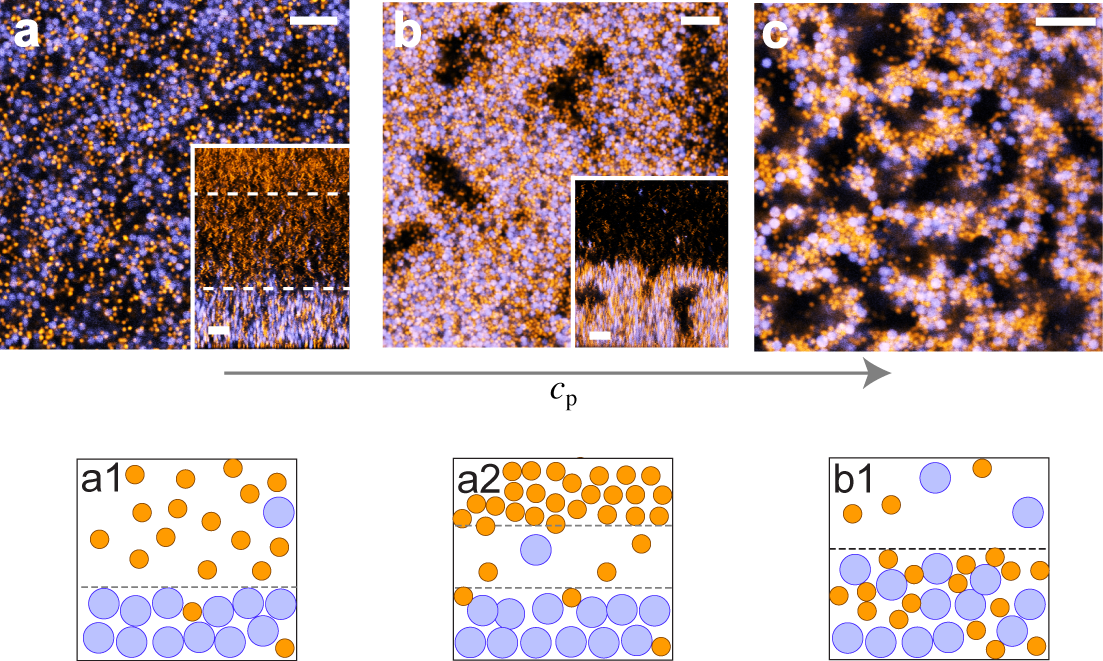}
\caption{Phase behaviour of binary colloid-polymer mixtures.
(\emph{a-c}) Confocal microscopy images (see text) with larger colloids (blue) and small (orange). Insets show $xz$ profiles which are 100 $\mu$m in height.  
(\emph{a}) $c_p/c_p^*=0.059$; Possible two- or three-phase demixing as indicated by dashed lines in inset. 
(\emph{b}) $c_p/c_p^*=0.069$; 
(\emph{c}) $c_p/c_p^*=0.090$, gelation.  Scale bars denote 10 $\mu$m. 
Possible scenarios by which the experimental data may be interpreted: 
(\emph{a1}) Two-phase coexistence --- the lower (liquid) phase is rich in large particles and the upper (vapor) phase is rich in small particles and polymer;
(\emph{a2}) The colloidal mixture exhibits liquid-liquid demixing. 
(\emph{b1}) Upon deeper quenching more small particles become entrained in the colloid-rich liquid.}
\label{figBinaryPhaseImages}
\end{figure*}

As the effective interactions between colloids resemble those found for binary mixtures of simple atomic fluids, we might expect to find phase behaviour similar to that proposed previously on the basis of mean field theories of a van der Waals mixture. Ref.~\cite{vankonyenburg1976} identified several classes of possible phase diagram topology which were subsequently found to apply to a wide range of real atomic and molecular mixtures \cite{rowlinson}. The classes are delineated by the degree of immiscibility of the two components. For mixtures in which the two species are not too dissimilar one of two scenarios is predicted: For type I behavior, the system exhibits only liquid-vapor phase separation, which for our case would correspond to coexistence between a colloid rich ``liquid'' phase (dilute in polymer) and a polymer rich ``vapor'' phase (dilute in colloids). Type II phase diagrams occur when the immiscibility is stronger: following liquid-vapor phase separation a further transition occurs at sufficiently large $c_p^r$ (low effective temperature) corresponding to a deep quench in which liquid-liquid demixing occurs at a critical end point, with a line of demixing critical points extending to higher densities. Below a critical end point, two liquids --- one rich in the larger colloids and the other in the small colloids --- coexist with a vapor phase. Possible scenarios are sketched in the second row of Fig. \ref{figBinaryPhaseImages}.

Although there are basic similarities with simple atomic mixtures, the colloidal system we study exhibits key differences: (\emph{i}) the ratio of the colloid diameters, 0.57, is rather large and (\emph{ii}) unlike atomic mixtures, the range of all three attractive (colloid-colloid) pair interactions is identical since this is set by the size of the single polymer species. Moreover the well depth of the effective depletion attraction between two large particles is about twice that between two small particles. Thus, it is not clear \emph{a-priori} which scenario for the fluid phase separation, type I or type II, should pertain in our system. It is also conceivable that, given that the interaction between the large colloids is stronger than that between the small, that the former might demix with the small acting in a similar way to a ``spectator phase''. Furthermore at sufficiently large $c_p^r$, colloids can undergo gelation,  which is not forseen in the classification scheme \cite{vankonyenburg1976}. Remarkably, we find that in our experiments the system appears to exhibit three-phase coexistence. However careful analysis informed by simulation reveals that this is illusory: there are two phases, yet their composition changes so drastically that it gives the impression of a new phase. We choose to term this non-monotonic behaviour of the ratio of volume fractions \emph{composition inversion.}

To understand this basic mixture, we combine particle-resolved experiments \cite{ivlev} with tailored Grand Canonical Ensemble (GCE) Monte Carlo (MC) simulation \cite{wilding2011}. Our experiments use confocal microscopy to provide real-space information on composition fluctuations and fractionation effects. These characterize phase coexistence and criticality in colloidal binary mixtures. Our system consists of two species, \emph{i.e.} two sizes of fluorescently labelled colloidal particles which are (nearly) density- and refractive index matched to their solvent. To this system polymer is added. Our simulations provide comparable information but are free from the influence of kinetics and gravity; they access equilibrium properties. In particular we obtain densities of coexisting phases and their spatial fluctuations. In our simulations, the effective colloid-colloid interactions are described by the Asakura-Oosawa (AO) model \cite{asakura1954,asakura1958}, generalised to a binary system.

This paper is organised as follows. In our methods section \ref{sectionMethods}, we discuss our experimental procedure in (section \ref{sectionExperimental}), the way in which we map our data between experiment and simulation is described in section \ref{sectionMapping}. The means we use to arrive at an effective Hamiltonian for the binary colloid system is described in section \ref{sectionEffective} and our tailored simulation methodology is introduced in section \ref{sectionTailored}.
In our results section \ref{sectionResults}, we describe the single-component colloid-polymer phase behaviour in section \ref{sectionPhaseBehaviorSingle} before proceeding to the phase behavior of the  binary colloids plus polymer which is the main experimental result of this work in section \ref{sectionPhaseBehaviorBinary}. Simulation results for the phase behaviour are presented in section \ref{sectionResultsFromSimulation}. Our finding of composition inversion is discussed in section \ref{sectionComposition}, and we complete our results section by showing the behaviour of near-critical fluctuations in section \ref{sectionNearCritical}. We conclude our paper in section \ref{sectionConclusions}.

\section{Methods}
\label{sectionMethods}

\subsection{Experimental details} 
\label{sectionExperimental}

%\textit{Sample preparation. --- }
Two sizes of Polymethylmethacrylate (PMMA) particles were used. Using static light scattering, the diameters of the large $l$ and small $s$ particles were determined to be $\sigma_l=1.84$ $\mu$m and $\sigma_s=1.04$ $\mu$m respectively, with polydispersity 5\% in each case. The colloid-colloid size ratio is then $q_{sl}=0.57$. The larger particles were labeled with the fluorescent dye 3,3-dioctadecyloxacarbocyanine perchlorate (DiOC$_{18}$), while the smaller particles were labeled with 1,1-dioctadecyl-3,3,3,3-tetramethylindocarbocyanine perchlorate (DiIC$_{18}$). The solvent used was a (near) refractive index- and density matched mixture of tetrachloroethylene and \emph{cis}-decalin. Polystyrene with molecular weight $m_w=1.5\times10^{7}$ g mol$^{-1}$ acted as the depletant, with radius of gyration $R_g\approx149$ nm in the good solvent used \cite{royall2007jcp}, giving polymer-colloid size ratios of $q_l=2R_g/\sigma_l=0.16$ and $q_s=2R_g/\sigma_s=0.29$ for the large and the small particles, respectively. The sizes of the particles and the depletant were chosen such that a stable colloidal liquid should exist for the mixture with only smaller particles while  in the case of the larger particles, the liquid is metastable to crystallization\cite{poon2002,lekkerkerker}.

We work in the vicinity of the critical isochore. For each pure colloid species $l$ and $s$ the critical colloid volume fraction was estimated from the literature\cite{loverso2006,fortini2008,taylor2012,royall2007} to be $\phi_l^c=0.26$ and $\phi_s^c=0.22$ respectively. For the mixture, an intermediate total overall volume fraction of $\phi_{tot}\equiv \phi_s+\phi_l=0.24$ was chosen, with $\phi_s=\phi_l$, implying that the overall concentration of large particles is $X_l\equiv N_l/(N_s+N_l)=0.143$. We focus on state points along this isochore distinguished by the choice of polymer mass fraction $c_p$. Since colloid-polymer mixtures have large critical regimes with relatively flat binodals, a precise determination of the critical isochore is not essential for the purposes of observing near-critical fluctuations \cite{royall2007}. When plotting experimental results, we use the ratio $c_p/c_p^\ast$ where $c_p^\ast$ is the value at overlap.

%\textit{Phase behavior and local volume fraction. --- } 
Phase diagrams in the experiments are determined as follows. In single phase regions (one-phase fluid or gel) we quote the colloid volume fraction at which the sample is prepared. In the case of phase separation, a sedimentation profile is used to determine the volume fraction of each colloid species as a function of height.

Sedimentation profiles were obtained from an intensity analysis which was calibrated using images of samples having known volume fraction. The fraction of the total intensity due to each species is nearly linearly dependent on the volume fraction of that species \cite{leocmach2010}. Given the calibration, we then determine the volume fraction as a function of position. Specifically, the volume fractions of both colloid species ${\phi_i(x,y,z)}$ where $i=l,s$ were calculated from an intensity average fraction around the point ${(x,y,z)}$ i.e.

\begin{equation}
\phi_i(x,y,z)\approx{C}\frac{\sum_{x=-L_x}^{L_x}\sum_{y=-L_y}^{Ly}\sum_{z=-L_z}^{L_z}I_i(x,y,z)}{8I_{max}N_xN_yN_z}
\label{eqI2phi}
\end{equation}

\noindent where ${C \approx 1\pm0.012}$ is a calibration constant, ${N_x}$,${N_y}$,${N_z}$  are the number of pixels in ${[-L_j,L_j]}$. In our case ${L_i}$ is selected to be $7.0{\sigma_l}$ as in the simulations and ${j\in\{x,y,z\}}$. $I_{max}=255$ is the maximum intensity, scaled by 8 to reflect the two channels. The sedimentation profile was obtained by scanning the ${xy}$ plane at every ${z}$ position.

\subsection{Mapping experiment to simulation}
\label{sectionMapping}

As noted in the introduction, our experimental results are plotted in terms of the dimensionless ratio $c_p/c_p ^\ast$ where $c_p^\ast$ is the polymer mass fraction at overlap and $c_p$ refers to the polymer mass fraction in the actual (experimental) polymer-colloid mixture. On the other hand, models and simulations of such mixtures are most naturally formulated in terms of a polymer reservoir, with a given chemical potential $\mu_p$, that is in osmotic equilibrium with the actual system. One can convert from the reservoir to the system representation if one knows the free volume fraction $\alpha(\phi_l,\phi_s;z_p)$ that relates the number density of polymers in the system to that in the reservoir $\rho_p^r$

\begin{equation}
   \rho_p(\phi_l,\phi_s;z_p) = \alpha(\phi_l,\phi_s;z_p) \rho_p^r(z_p)        
   \label{eqRho}  
\end{equation}

\noindent where $z_p$ is the fugacity of the polymer. Free volume arguments suggest that the free volume fraction can be approximated by its value in the limit $z_p \to 0$, i.e. vanishing polymer density. Within the context of the Asakura-Oosawa
model, where the polymer is ideal and $z_p= \rho_p^r$ (see next subsection), the free volume fraction is easily calculated from Percus-Yevick (PY) results, equivalent to scaled particle theory, for the excess chemical potential of a binary HS mixture \cite{lekkerkerker1992,dijkstra1999}. We have generalized this approach to the ternary HS case and within the PY approximation we find the following result in the limit $z_p \to 0$ :

\begin{equation}
\alpha^\mathrm{PY}(\phi_l,\phi_s;z_p=0)=(1-\phi_\mathrm{tot}) \mathrm{exp}\left[  -\tilde{A} \tilde{\gamma}     -\tilde{B} \tilde{\gamma}^2     -\tilde{C} \tilde{\gamma}^3  \right]
\label{eqAlphaALaBob}
\end{equation}

\noindent
where

\begin{equation}
\tilde{\gamma} = \frac{1}{1-\phi_\mathrm{tot}},
\label{eqGamma}
\end{equation}

\begin{equation}
\tilde{A}=\phi_l q_l^3 + \phi_s q_s^3 + 3(\phi_l q_l^2 + \phi_s q_s^2) + 3( \phi_l q_l+ \phi_s q_s),
\label{eqA}
\end{equation}

\begin{equation}
\tilde{B}=\frac{9}{2}(\phi_l q_l + \phi_s q_s)^2 + 3(\phi_l q_l^2 + \phi_s q_s^2)( \phi_l q_l+ \phi_s q_s),
\label{eqB}
\end{equation}
and

\begin{equation}
\tilde{C}=3(\phi_l q_l + \phi_s q_s)^3.
\label{eqC}
\end{equation}

In the limit where the volume fraction of one colloid species vanishes this result reduces to that of Lekkerkerker et al
\cite{lekkerkerker1992}. For our experimental conditions, $q_l= 0.16, q_s=0.29, \phi_l=\phi_s =0.12$ we find that this approximation gives a free volume fraction $\alpha^{PY}\sim 0.6$. Theory and simulation for the AO model usually work with the polymer reservoir volume fraction $\phi_p^r =\pi\sigma_p^3\rho_p^r /6$. This quantity sets the strength of the attractive interactions-see Eq. (\ref{eqUlseff}) and Fig.~\ref{figAOpots}. For example, our simulations yield a critical point at $\phi_p^r = 0.375(5)$. In the text we use the term $c_p^r$, the polymer reservoir concentration, to denote $\phi_p^r$ . Since for fixed $\phi_l$ and $\phi_s, \alpha^{PY}$ in (\ref{eqAlphaALaBob}) is constant it follows that that to a good approximation we can convert from simulation to experiment assuming $c_p/c_p^\ast= {\rm const.} \phi_p^r$. We fix the constant by matching the critical points in simulation and experiment. We estimate the experimental critical point to be at $c_p/c^*_p = 0.057 \pm 0.002$ -- see Fig.~\ref{figInversionPhaseNigel}(\emph{iii}). There is a small deviation from linearity ($<5\%$) upon phase separation. Although we could correct for this using the appropriate colloid volume fractions in $\alpha^{PY}$ this would not remove other errors in the mapping. These arise from polymer non-ideality, deformation and other deviations from the ideal AO model \cite{bolhius2002,aarts2002,louis2002jcp,krakoviack2003}.

\subsection{The effective two-component Hamiltonian.}
\label{sectionEffective}

In this subsection, we describe the model that we investigate in simulations. We consider a ternary system consisting of two species of colloids, modelled as large and small hard-spheres (HS) with different diameters $\sigma_l$, $\sigma_s$, plus a single polymer species $p$. The Hamiltonian is

\begin{equation}
H=H_{ll}+H_{ss}+H_{ls}+H_{lp}+H_{sp}+H_{pp}\:,
\label{eqHam}
\end{equation}
where $H_{ll}$ denotes hard sphere (HS) interactions between $ll$, $H_{ss}$ denotes HS interactions between $ss$ and $H_{ls}$ those between unlike species. The $ls$ HS interaction potential $u_{ls}^{HS}(r)$ is assumed additive so that the 
cross-diameter $\sigma_{ls}\equiv(\sigma_l+\sigma_s)/2$. The polymer coils are treated as mutually interpenetrable 
(non-interacting or ideal) so that $H_{pp}=0$. However, the centre of mass of a coil is excluded from the large colloid centre to a distance $(\sigma_l+\sigma_p)/2$ or $(\sigma_s+\sigma_p)/2$ for the small colloid. The diameter of the polymer is $\sigma_p=2R_g$, where $R_g$ is the radius of gyration of the polymer. Equation~\ref{eqHam} defines the Asakura-Oosawa
(AO) model for our present ternary mixture \cite{asakura1954,asakura1958,vrij1976}. Henceforward we ignore trivial kinetic energy terms.

Following \cite{dijkstra1999} we work in the semi-grand ensemble where the numbers $N_l$ and $N_s$ of the large and small HS are fixed, as are the volume $V$, inverse temperature $\beta$ and the polymer fugacity $z_p=\Lambda_p^{-3}\exp(\beta\mu_p)$.  Here $\Lambda_p$ is the thermal de Broglie wavelength and $\mu_p$ is the chemical potential of the polymer reservoir. For ideal polymer, we recall $z_p=\rho_p^r$, the polymer density in the reservoir. The thermodynamic potential $F$ appropriate to this ensemble is given by a direct generalization of Eq (3) in \cite{dijkstra1999} and the effective Hamiltonian of the two-component colloid mixture, obtained by integrating out the polymer degrees of
freedom, takes the form

\begin{equation}
H^\mathrm{eff}=H_{ll}+H_{ss}+H_{ls}+\Omega\:,
\label{eqHeff1}
\end{equation}
where $\Omega$ is the grand potential of the fluid of ideal polymer in the field of a fixed configuration of $N_l$ and $N_s$ HS colloids; $\Omega$  depends on the coordinates of both HS species \cite{dijkstra1999}.

Extending the analysis presented in \cite{dijkstra1999} to the binary HS case leads directly to a diagrammatic expansion of $\Omega$ that generalizes Eq (6) of \cite{dijkstra1999}, i.e. $\Omega$  is a sum of zero, one-body, two-body and higher-body colloidal terms that involve integrals over products of $lp$ and $sp$ Mayer bonds. The upshot is that the effective Hamiltonian takes the form:

\begin{widetext}
\begin{equation}
H^\mathrm{eff}= H_{0}+\sum_{i<j}^{N_l}u_{ll}^\mathrm{eff}(R_{ij})+\sum_{i<j}^{N_s}u_{ss}^\mathrm{eff}(R_{ij})
           +\sum_{i=1}^{N_l}\sum_{j=1}^{N_s}u_{ls}^\mathrm{eff}(R_{ij})+{\rm H.O.~~terms}
\label{eqHeff2}
\end{equation}  
\end{widetext}
where $R_{ij}$ is the distance between the centres of particles $i$ and $j$ and the effective {\it ll} (or {\it ss}) pair potential $u^\mathrm{eff}$ is that pertaining to a one-component HS {\it l} (or {\em s}) AO system with the appropriate HS diameter $\sigma_l$ (or $\sigma_s$) \cite{dijkstra1999}. The new two-body term is the effective $ls$ pair potential which we write out explicitly below. The first term in Eq.~(\ref{eqHeff2}) is the sum of zero and one-body terms which, for a uniform mixture with constant densities, is

\begin{figure}[!htb]
\centering
\includegraphics[width=40mm]{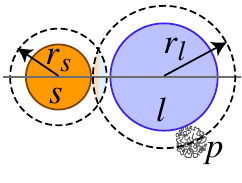}
\caption{Geometry for depletion interaction between unlike colloids {\em l} and {\em s} due to polymer {\em p}. The radius of the depletion sphere around {\em l} is $(\sigma_l+\sigma_p)/2$ and that around {\em s} is $(\sigma_s+\sigma_p)/2$. The overlap lens shape is indicated; its volume determines the depletion potential given in Eq.~(\ref{eqUlseff}).}
\label{figDougPaddy}
\end{figure}

\begin{equation}
\beta H_0=-z_pV\left[1-\phi_l(1+q_l)^3-\phi_s(1+q_s)^3\right],
\end{equation}

\noindent
where $\phi_l=\pi\rho_l\sigma_l^3/6$, with number density $\rho_l=N_l/V$, is the volume fraction of the large ($l$) HS and equivalently for $s$. As noted earlier, the size ratios are $q_l=\sigma_p/\sigma_l$, $q_s=\sigma_p/\sigma_s$. Since $H_0/V$ depends linearly on $\rho_l$ and $\rho_s$, this term does not affect the phase equilibria \cite{dijkstra1999} which is the concern of the present study. The higher order terms in (\ref{eqHeff2}) correspond to 3-body, 4-body etc. effective inter-colloidal interactions. Generally these terms are non-zero and, as the size ratios increase, we expect an increasing number of higher-body contributions. However, for a sufficiently asymmetric case, i.e. with $q_s<(2/\sqrt{3}-1)= 0.1547$ and following arguments of \cite{dijkstra1999}, it is easy to show that three and higher-body terms vanish identically in (\ref{eqHeff2}). Thus in this regime pair potentials alone determine phase equilibria. This is an important result. It implies that for sufficiently asymmetric cases, the ternary AO system can be mapped {\em exactly} to a two-component colloid mixture in which the colloids interact solely through pair potentials. 
We return to this observation below.

\begin{figure}[!htb]
\centering
\includegraphics[width=75mm]{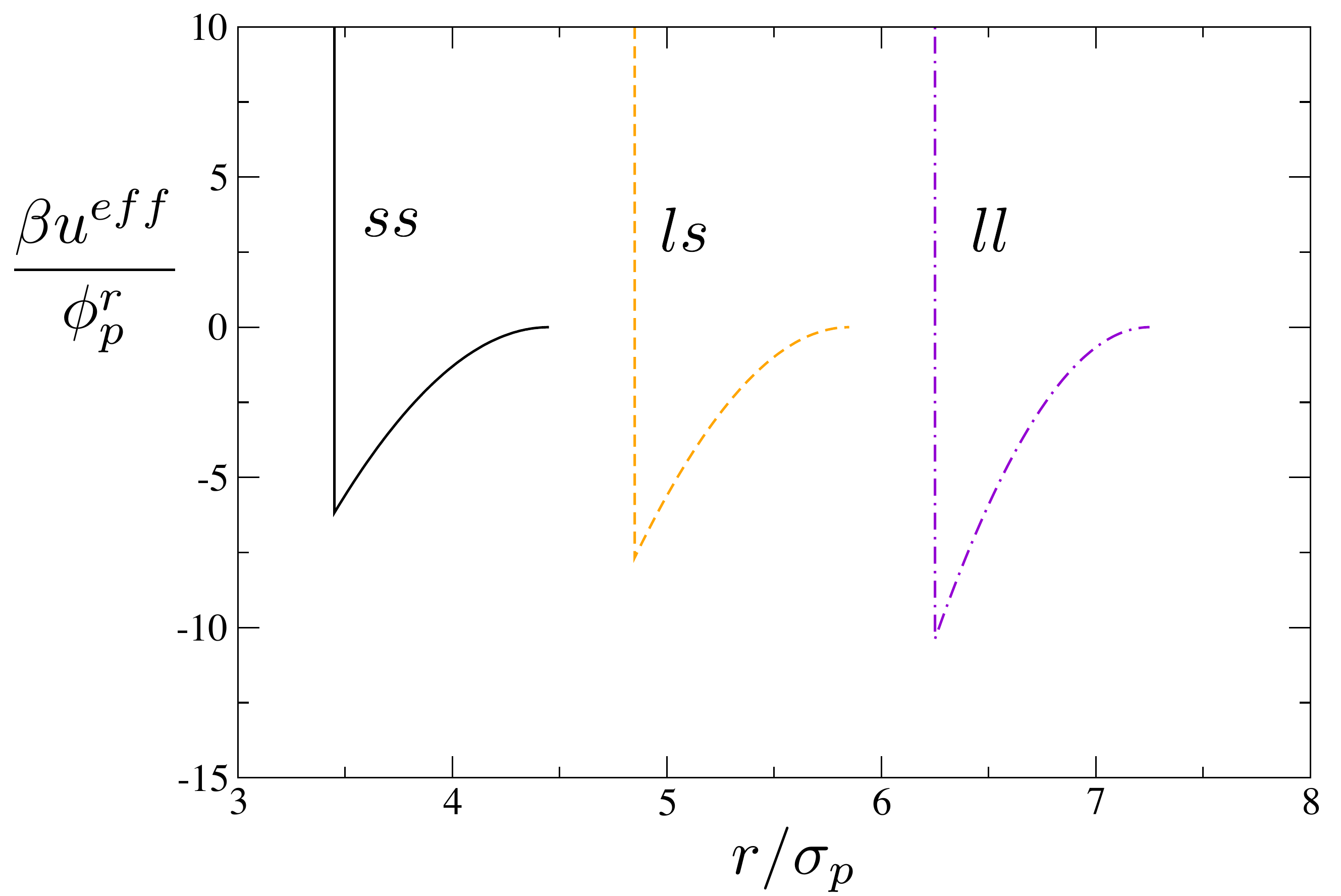}
\caption{The three effective pair potentials $\beta u^{eff}(r)$ plotted versus $r/\sigma_p$ for size ratios $q_l =0.16$ and $q_s=0.29$. Note that the range of the attraction is $\sigma_p$ for all three potentials.}
\label{figAOpots}
\end{figure}

It remains to specify the $ls$ effective pair potential. This is easily calculated. The attractive depletion or AO potential  $u_{ls}^{AO}(r)$ is equal to the volume of the lens formed by the overlap of depletion layers around $l$ and $s$ times $-z_p\beta^{-1}$. The geometry of the overlap is illustrated in Fig.~\ref{figDougPaddy}. We find:

\begin{widetext}
\begin{eqnarray}
\beta u_{ls}^\mathrm{eff}(r)&=&\beta u_{ls}^{HS}(r)+ \beta u_{ls}^{AO}(r)\nonumber\\
&=& \left \{ \begin{array}{ll} 
      \infty \:,  \hfill 0<r < \sigma_{ls} & \\[1mm]
      \frac{\phi_p^r}{\sigma_p^3}    \frac{(\sigma_{ls}+\sigma_p-r)^2[3(\sigma_l-\sigma_s)^2-8r(\sigma_{ls}+\sigma_p)-4r^2]}{8r} \:, 
      \mbox{\hspace{2cm}}  \hfill\sigma_{ls}<r < \sigma_{ls} +\sigma_p    &       \\[2mm]
      0 \:, \hfill r > \sigma_{ls} +\sigma_p    & 
\end{array}\right.\nonumber\\
\label{eqUlseff}
\end{eqnarray}
\end{widetext}
where $r$ is the distance between the centres of colloid $l$ and colloid $s$ and $\phi_p^r=\pi\rho_p^r\sigma_p^3/6$ is the volume fraction of polymer in the reservoir. It is straightforward to show that (\ref{eqUlseff}) reduces to the standard one-component AO result when $\sigma_l=\sigma_s=\sigma_{ls}$ \cite{dijkstra1999}. The three effective pair potentials $u_{ll}^{eff}(r)$, $u_{ss}^{eff}(r)$, $u_{ls}^{eff}(r)$ have different hard-core diameters but exhibit identical finite range of attraction, equal to $\sigma_p$, the diameter of the single polymer species. These pair
potentials, employed in our computer simulations, are each proportional to $\phi_p^r$ which implies that this quantity plays the same role as does inverse temperature in simple atomic fluids. A plot of the potentials, divided by $\phi_p^r$, is given in Fig.~\ref{figAOpots} for the experimental size ratios $q_l =0.16$ and $q_s =0.29$. For these parameters the depth of the $ll$ depletion potential is about $1.7$ times the $ss$ depth while the $ls$ depth is about $1.24$ times the $ss$ depth. Note that these pair potentials are somewhat different from those one might choose to model a binary mixture of atomic fluids, say Xe and Ar. In our case, the range of the interaction is identical for all these pair potentials wheras for the atomic case the range increases with the size and polarizability of the species \cite{rowlinson}.

Although the value of $q_l$ we employ is only very slightly greater than $0.1547$, \emph{i.e.} the value where three-body contributions begin to contribute, $q_s$ is considerably larger. This implies that in mapping the ternary AO model, for these particular parameters, to the effective two-component mixture some many-body interactions are omitted. One can estimate the importance of the latter by considering the mapping of the standard AO model with species $s$ only. For $q_s =0.29$ the pair potential description provides an accurate description for the full binary AO mixture \cite{taffs2010}.

\subsection{Tailored simulation methods}
\label{sectionTailored}

We employ grand canonical ensemble (GCE) Monte Carlo simulation to study a binary mixture of particles interacting via the AO potential of $u^{eff}_{ll},u^{eff}_{ss}$ and $u^{eff}_{ls}$ entering Eq. ~\ref{eqHeff2} %Eq.~\ref{eqUlseff}, 
with $q_l=0.16$ and $q_s=0.29$. Use of the GCE allows accurate and efficient simulation of fluid phase transitions and critical phenomena because it provides for density fluctuations on the scale of the simulation box. Traditional approaches of MD simulation in the microcanonical or canonical ensembles, whilst more straightforward to implement, lead to accuracy problems and, particularly for fluid mixtures, to enhanced finite-size effects \cite{wilding2010jcp}. Our approach is tailored to exploit the accuracy and flexibility of the GCE whilst simultaneously adhering to the experimental conditions of fixed overall volume fractions of the two components: $\phi_l=\phi_s=0.12$. The challenge is to satisfy this global constraint on average, even when the system has separated into $m$ coexisting phases, each occupying a certain proportion of the total volume. Under these conditions the
coexisting phases are generally ``fractionated'', \emph{i.e.}. their compositions differ from one another and one should like to determine the composition of each phase and the fraction of the system that it occupies.

To see how this can be achieved, consider the distribution of particles between the phases. This is described by a generalized lever rule:

\begin{eqnarray}
\label{eq:levera}
\phi_l^{(0)}=\sum_{\gamma=1}^m \xi^{(\gamma)}\phi_l^{(\gamma)}\:,\\ %\addtocounter{abc}{1}
%\addtocounter{equation}{-1}
\phi_s^{(0)}=\sum_{\gamma=1}^m \xi^{(\gamma)}\phi_s^{(\gamma)}\:.
\label{eq:leverb}
\end{eqnarray}
Here $\phi_l^{(0)}=N_l\sigma_l^3/(6V), \phi_s^{(0)}=N_s\sigma_s^3/(6V)$ are the overall (global) volume fractions of the two species {\em l} and {\em s}; $\xi^{(\gamma)}$, $\gamma=1,\ldots, m$ is the phase fraction of phase $\gamma$ which satisfies $\sum_{\gamma=1}^m\xi^{(\gamma)}=1$; $\phi_l^{(\gamma)}, \phi_s^{(\gamma)}$ are the volume fractions of the individual components in phase $\gamma$. It follows that in order to specify the coexistence properties of the system for some prescribed $\phi_s^{(0)}$ and $\phi_l^{(0)}$,  one must determine $\xi^{(\gamma)},\phi_s^{(\gamma)},\phi_l^{(\gamma)}$ for each phase $\gamma$. This can be done iteratively within a histogram reweighting framework, using a variant of an approach originally developed in the context of polydisperse fluids \cite{buzzacchi2006}. Specifically, for given $\phi_s^{(0)}, \phi_l^{(0)}$ and $\phi_p^r$, one regards the chemical potentials $\mu_s$ and $\mu_l$, and the phase fractions $\xi^{(\gamma)}$ as parameters to be tuned such as to satisfy both the generalized lever rule Eq.~(\ref{eq:levera}, \ref{eq:leverb}) and equality of the probability of the phases. For this purpose it is expedient to define a suitable order parameter probability distribution function (pdf), such as a density or composition distribution, which exhibits distinct peaks, one for each phase. The equality of the peak weights determines the conditions for which the phases are equally probable, which within the GCE implies the phases have equal pressure. Additionally, the peaks in the pdf allow one to assign any given configuration to a phase on the basis of its order parameter. This in turn permits the ready determination of the ensemble-averaged volume fractions $\phi_s^{(\gamma)}$ and $\phi_l^{(\gamma)}$ which appear in the lever rule. Since the order parameter pdf typically exhibits large probability barriers corresponding to mixed phase states, its form is best determined using multicanonical preweighting \cite{wilding2010jcp,wilding2011}.

Use of this method allows $\xi^{(\gamma)},\phi_s^{(\gamma)}, \phi_l^{(\gamma)}$ to be determined with finite-size errors which are exponentially small in the system size \cite{buzzacchi2006}. This is true even if the prescribed coexistence state point lies close to the phase boundary, ie. close to one end of a coexistence tie line, where the phase fraction of one phase vanishes. Standard methods for determining phase coexistence properties struggle in this regime because the minority phase contains very few particles. In our method however, the phases that occur near the end of the coexistence tie line are instead studied under conditions corresponding to the center of the tie line. Here the system fluctuates with equal probability between configurations in which each phase fills the simulation box in turn. This minimizes finite-size effects, while application of the lever rule condition allows us to infer accurately the phase fractions corresponding to the state point of interest close to the phase boundary.

\section{Results}
\label{sectionResults}

\subsection{Phase Behavior of single colloid species-polymer mixtures: Experiment} 
\label{sectionPhaseBehaviorSingle}

\begin{figure}[!htb]
\centering
\includegraphics[width=60mm]{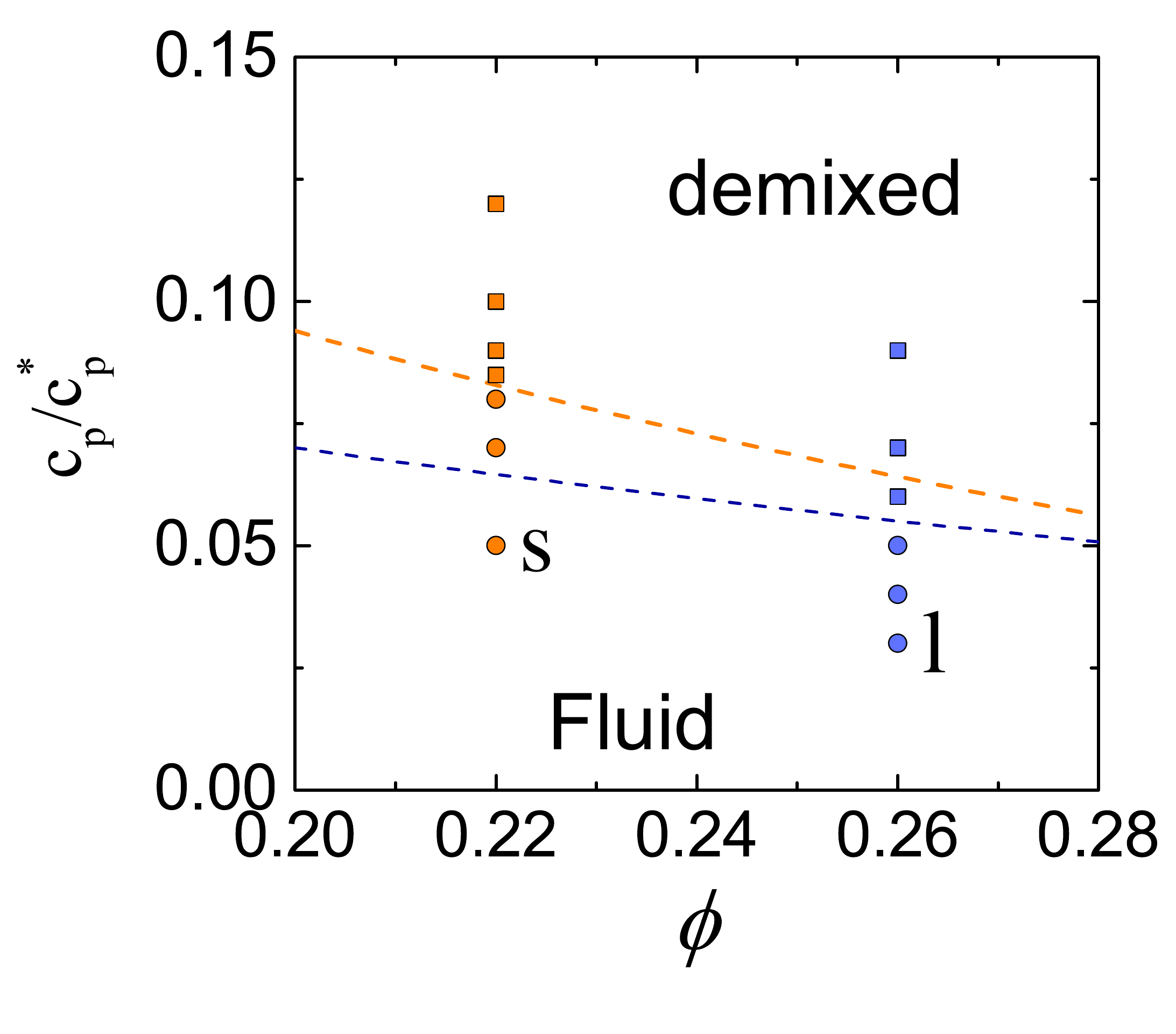}
\caption{Phase diagrams of colloid-polymer mixtures with one species of colloids.  Orange data points  are for the small particles ($q_s=0.29$), with critical volume fraction $\phi^c_s=0.22$. Blue data points  are for the larger particles ($q_l=0.16$), with critical volume fraction $\phi^c_l=0.26$.  Squares indicate phase coexistence or gelation; circles indicate a one-phase fluid.  The dashed lines indicate purported phase boundaries ascertained from confocal images taken at various state points.
}
\label{figSinglePhaseDiagram}
\end{figure}

We begin by noting the phase behavior of mixtures consisting of a single colloid species and polymer. For a sample comprised solely of large particles at the estimated critical colloid volume fraction of $\phi_l^c = 0.26$, liquid-vapor phase separation occurs at $c_p^c/c_p^\ast = 0.055\pm 0.005$; likewise, for a sample comprised solely of small particles at the estimated critical volume fraction of $\phi_s^c=0.22$, phase separation occurs at $c_p^c/c_p^\ast = 0.0825\pm 0.0025$. The phase boundaries for both systems are indicated in Fig. \ref{figSinglePhaseDiagram}.%S1 of the SI. 

\subsection{Phase behavior of binary colloid-polymer mixtures: Experiment} 
\label{sectionPhaseBehaviorBinary}

The bidisperse colloid mixture undergoes phase separation at $c_p^c/c_p^\ast = 0.057\pm 0.002$, indistinguishable from the system of large particles only. Confocal images in the $xy$ plane at height $z$ near the bottom of the container are shown in Fig.~\ref{figBinaryPhaseImages}. The insets show the system in the $xz$ plane. Consider first Fig.~\ref{figBinaryPhaseImages}(\emph{a}) 
which is for $c_p/c_p^\ast = 0.059$, corresponding to a shallow quench to a state point just within the phase coexistence region. The inset shows separation into two (or possibly three --- see later) phases, the denser of which has sedimented. The upper phase is overwhelmingly composed of small particles (orange) while the lower phase contains the vast majority of the large particles (blue). However, the main panel reveals substantial numbers of small particles in the dense phase, as well as significant spatial density fluctuations. On performing a deeper quench to $c_p/c_p^\ast=0.069$ [Fig.~\ref{figBinaryPhaseImages}(\emph{b})], one finds very different structure. From the inset,  one observes phase separation with the denser phase sedimenting, but now there are many more small particles in the lower phase and very few in the upper phase. Finally upon further quenching to $c_p/c_p^\ast=0.090$ [Fig. \ref{figBinaryPhaseImages}(\emph{c})], the system undergoes dynamical arrest and a gel forms. Both species occupy the dense interpenetrating arms of the gel. A schematic phase diagram based on analysis of these images is given in Fig.~\ref{figInversionPhaseNigel}(\emph{iii}).

Despite the particle-level detail, the results of Fig.~\ref{figBinaryPhaseImages} do not readily permit one to distinguish between type I and II phase behaviour. Specifically, Fig.~\ref{figBinaryPhaseImages}(\emph{a}) could be interpreted in a number of ways. The observation that the small colloids are fairly uniformly distributed among the phases could be taken to imply that the large colloids are somehow behaving as an effective one-component system which has undergone liquid-vapor phase separation, while the small particles only ``spectate'' in this process. Alternatively, it might be more appropriate to think in terms of the mixture as a whole undergoing liquid-vapor phase separation, but with a strong fractionation of the large particles to the liquid phase and only weak fractionation of the small particles. This scenario is depicted schematically in Fig.~\ref{figBinaryPhaseImages}(\emph{a1}). A further possible interpretation of Fig.~\ref{figBinaryPhaseImages}(\emph{a}) is that liquid-liquid demixing occured [Fig.~\ref{figBinaryPhaseImages}(\emph{a2})] and two colloidal liquid phases coexist with the third, polymer-rich colloidal vapor, as indicated by the dashed lines in Fig. \ref{figBinaryPhaseImages}(\emph{a}) inset. However, if such type II behaviour occurs, it is curious that the small particles subsequently remix with the large ones in a dense phase at larger polymer concentration $c_p/c_p^\star$ as seen in Fig.~\ref{figBinaryPhaseImages}(\emph{b}),(\emph{b1}).
%(\emph{b1})

\subsection{Results from Simulation}
\label{sectionResultsFromSimulation}

\begin{figure*}%[!htb]
\centering
\includegraphics[width=180mm]{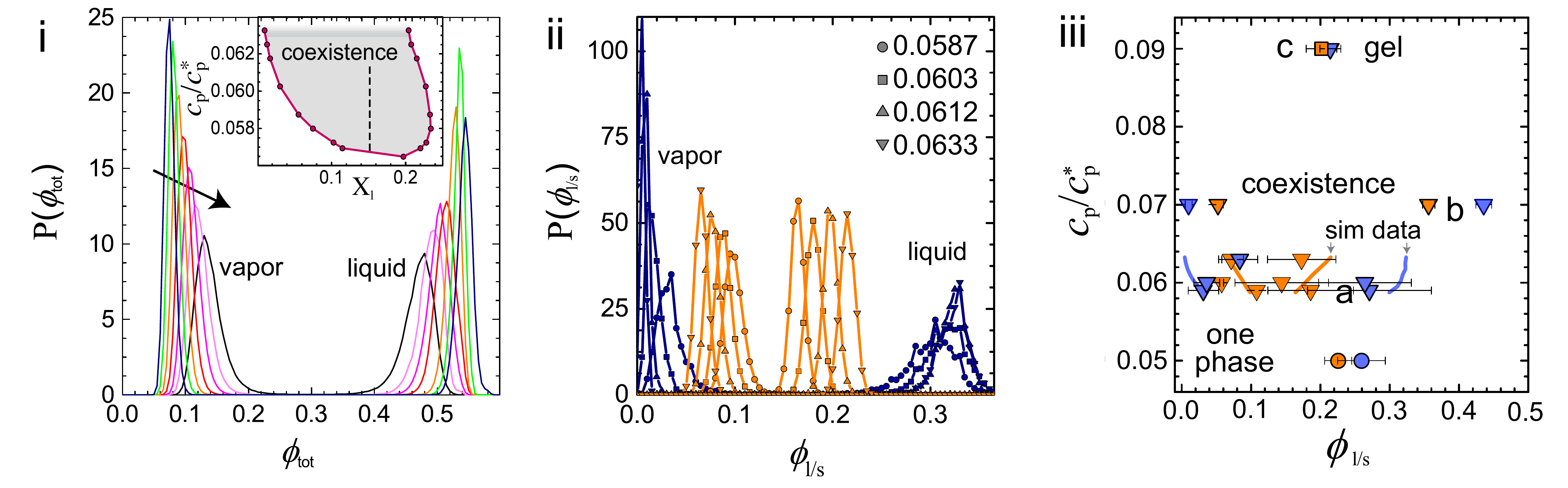}
\caption{
(\emph{i}) Grand Canonical Monte Carlo simulation results for the probability distribution $P(\phi_\mathrm{tot})$ in the vapor and liquid phases for various polymer reservoir concentrations $c_p^r$. The arrow denotes reducing $c_p^r$ and mapping to $c_p/c_p^*=0.0633$ to $0.0587$ in equal increments. The inset shows the large particle concentration $X_l$ in the vapor and liquid phases with the overall composition $X_l=0.143$ marked as a dashed vertical line; Note the maximum of $X_l$ in the liquid phase denotes composition inversion as detailed in the text.
(\emph{ii}) Simulation results for $P(\phi_l)$ and $P(\phi_s)$ in the vapor and liquid phases for values of $c_p/c_p^\ast$ shown in the key. Orange data denote small particles and blue large.
(\emph{iii})  Schematic phase diagram. 
(a,b,c) refer to state points obtained in experiment depicted in Fig. \ref{figBinaryPhaseImages}.
Squares are gels, triangles are liquid-vapour coexistence, circles are one phase fluid.
Simulation results for volume fractions of coexisting phases are given by pale blue and orange lines.
}
\label{figInversionPhaseNigel}
\end{figure*}

To help resolve which scenario applies, we appeal to our Grand Canonical Monte Carlo simulation studies of the generalized Asakura-Oosawa model. Starting from the one phase regime, the polymer reservoir concentration $c_p^r$ was increased following the experimental isochore until the systems entered the coexistence region. This is indicated by the appearance of a double peaked structure in the probability distribution of the fluctuating order parameter (which we take as the total volume fraction $\phi_{tot}=\phi_l+\phi_s$), as shown in Fig. \ref{figInversionPhaseNigel}(\emph{i}). One of these peaks is at very low values of $\phi_{tot}$, while the other is at a high value, indicating that the transition is vapor-liquid like in character. We have followed the transition to large $c_p^r$ where the liquid becomes very dense, but see no sign of a splitting of the liquid peak that would indicate liquid-liquid demixing \emph{i.e.} type II behavior. At higher densities, it becomes difficult to sample the liquid sufficiently in our simulations.

Thus, the simulations indicate that only a single vapor-liquid transition occurs implying type I behavior. Moreover they reveal
that the puzzling differences between Fig.~\ref{figBinaryPhaseImages}(\emph{a}) and (\emph{b}) (which suggests possible liquid-liquid demixing of colloids) might be attributed to the changing character of the fractionation as $c_p^r$ is varied. Figure~\ref{figInversionPhaseNigel}(\emph{b}) plots the probability distributions $P(\phi_s)$ and $P(\phi_l)$ of the volume fractions of each species in the coexisting vapor and liquid phases for the various $c_p^r$ studied. One observes from these distributions that the volume fraction difference of the large particles $\phi_l^\mathrm{liq}-\phi_l^\mathrm{vap}$ is very large even for small $c_p^r$ approaching the critical point. This indicates that the vast majority of large particles occupy the liquid from the outset of phase separation. On increasing $c_p^r$ this difference grows further until, at the largest $c_p^r$ studied, almost no large particles occupy the vapor. With regard to the small particles, at low values of $c_p^r$, $\phi_s^\mathrm{liq}$ exceeds $\phi_s^\mathrm{vap}$, only slightly, \emph{i.e.} there is initially only weak fractionation of the small particles upon phase separation. However as $c_p^r$ increases, $\phi_s^\mathrm{liq}-\phi_s^\mathrm{vap}$ grows strongly, indicating that the small particles migrate progressively from the vapor to the liquid. Figure~\ref{figInversionPhaseNigel}(\emph{iii}) summarises the phase behaviour as determined by experiment and simulation. Overall there is good agreement.

\subsection{Composition inversion} 
\label{sectionComposition}

An interesting corollary of the fractionation behavior is that the concentration of large particles $X_l$ in the liquid phase exhibits an unusual back-bending, \emph{i.e.} as $c_p^r$ increases a maximum occurs as $X_l$ increases [see the phase diagram in the inset of Fig.~\ref{figInversionPhaseNigel}(\emph{i})]. We term this behavior \emph{composition inversion}. It appears not to have been recognized previously in studies of binary mixtures.

\begin{figure*}%[!htb]
\centering
\includegraphics[width=120mm]{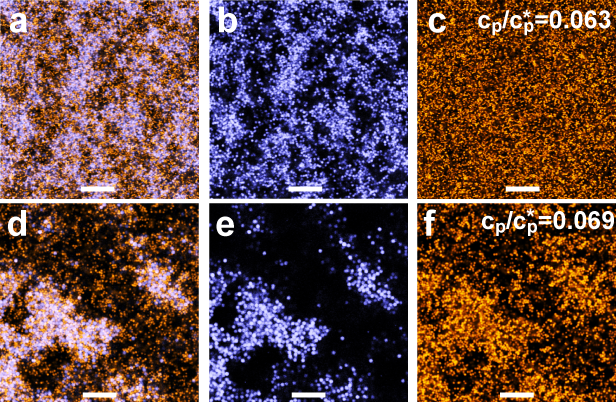}
\caption{Confocal images shortly after commencement of phase separation. The mixture [left panels (\emph{a}), \emph{(d}] is separated into the contributions from large [blue, middle panel (\emph{b}), \emph{(e)}] and small particles [orange, right panel (\emph{c}), \emph{(f)}]. \emph{(a-c)}, top row $c_p/c_p^\ast=0.063$; \emph{(d-f)}, bottom row) $c_p/c_p^\ast=0.069$. Scale bars denote 25 $\mu$m.}
\label{figChannels}
\end{figure*}

The fractionation scenario revealed by the simulations can explain the differences in the images of Fig.~\ref{figBinaryPhaseImages}\emph{(a,b)}. Figure~\ref{figChannels}(\emph{a}) shows that for weak quenching and early times (before sedimentation), the large particles accumulate in the liquid phase while the smaller are more homogeneously distributed. This can be seen by separating the fluorescent channels to reveal the spatial distributions of the individual species [Fig. \ref{figChannels}(\emph{b,c})]. At larger quench depths, Fig. \ref{figChannels}(\emph{d-f}), small particles follow the large particles in their spatial variation in density. In other words, the liquid phase is rich in both colloid species.

\subsection{Near-critical fluctuations} 
\label{sectionNearCritical}

At the vapor-liquid critical point one expects the system to display self-similar spatial density fluctuations on all length scales. By reference to our simulations, the experimental path enters the coexistence region slightly on the vapor side of the critical point. The presence of density fluctuations on many lengthscales, as observed in Fig.~\ref{figChannels} for a state point just inside the coexistence region, is therefore a reflection of the proximity to criticality. However, fractionation also affects the near-critical region: Principally it is the large particles that partake in these fluctuations -- the small ones are more homogeneously distributed. We have quantified this effect in both the experiments and simulations by accumulating the probability distributions $P(\phi_s)$ and $P(\phi_l)$.

The analysis of volume fraction fluctuations for the individual species in the experiments was obtained by plotting a histogram of the volume fractions obtained via (\ref{eqI2phi}), sampled over square regions of side $7.0\sigma_l$. This differs from the simulation analysis which obtains the distribution of the fluctuating species volume fractions on the scale of the cubic simulation box. Because of the limited axial resolution of the microscope, we do not define 3d cubes very accurately at this lengthscale.

\begin{figure}
\centering
\includegraphics[width=60mm]{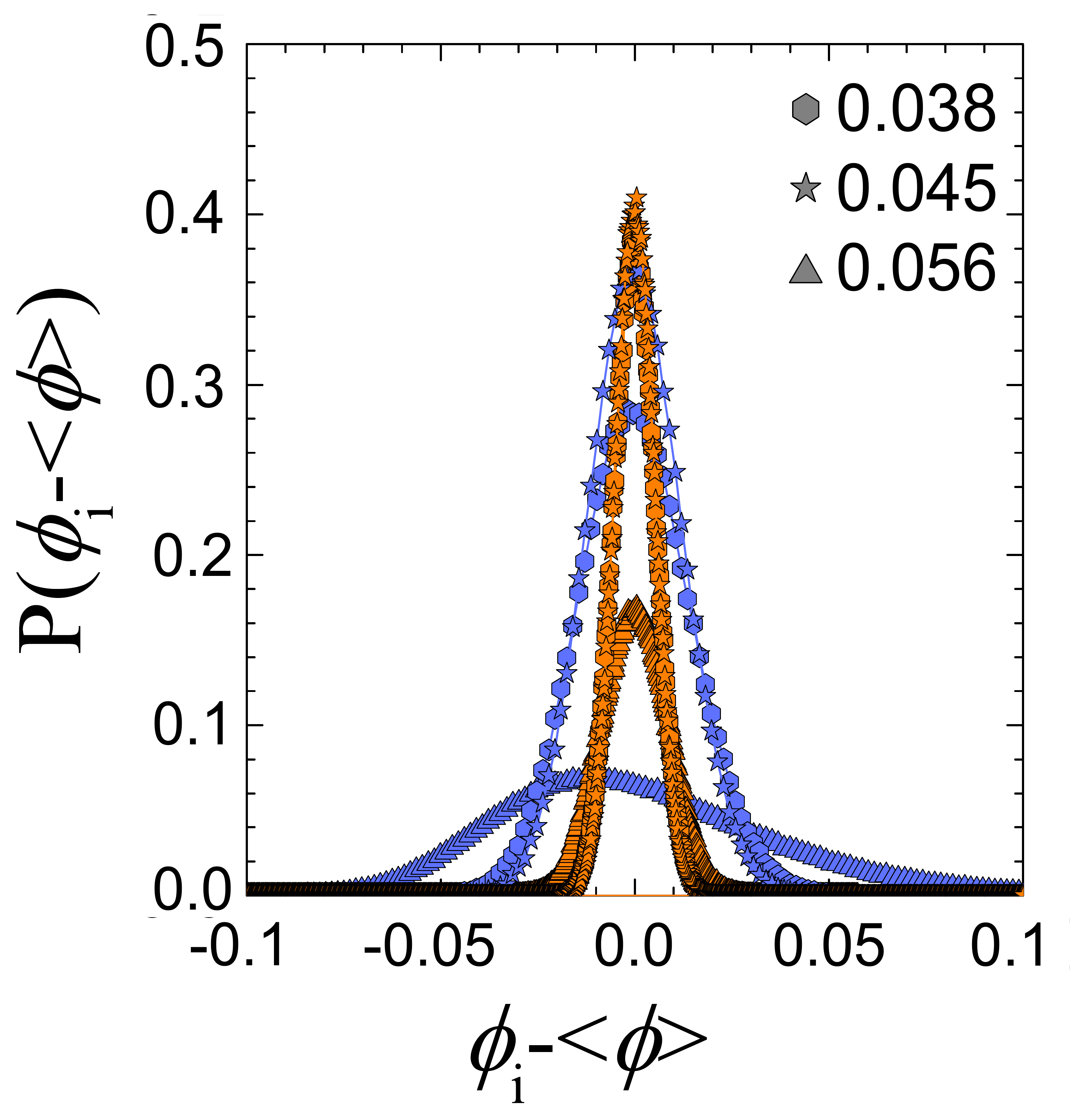}
\caption{Simulation results for $P(\phi_l)$ and $P(\phi_s)$ in the one phase fluid on the experimental isochore for three values of $c_p/c_p^\ast$. Orange data denote small particles $i=s$ and blue large $i=l$.}
\label{figIntensityFluctuationsSimulation}
\end{figure}

\begin{figure}
\centering
\includegraphics[width=55mm]{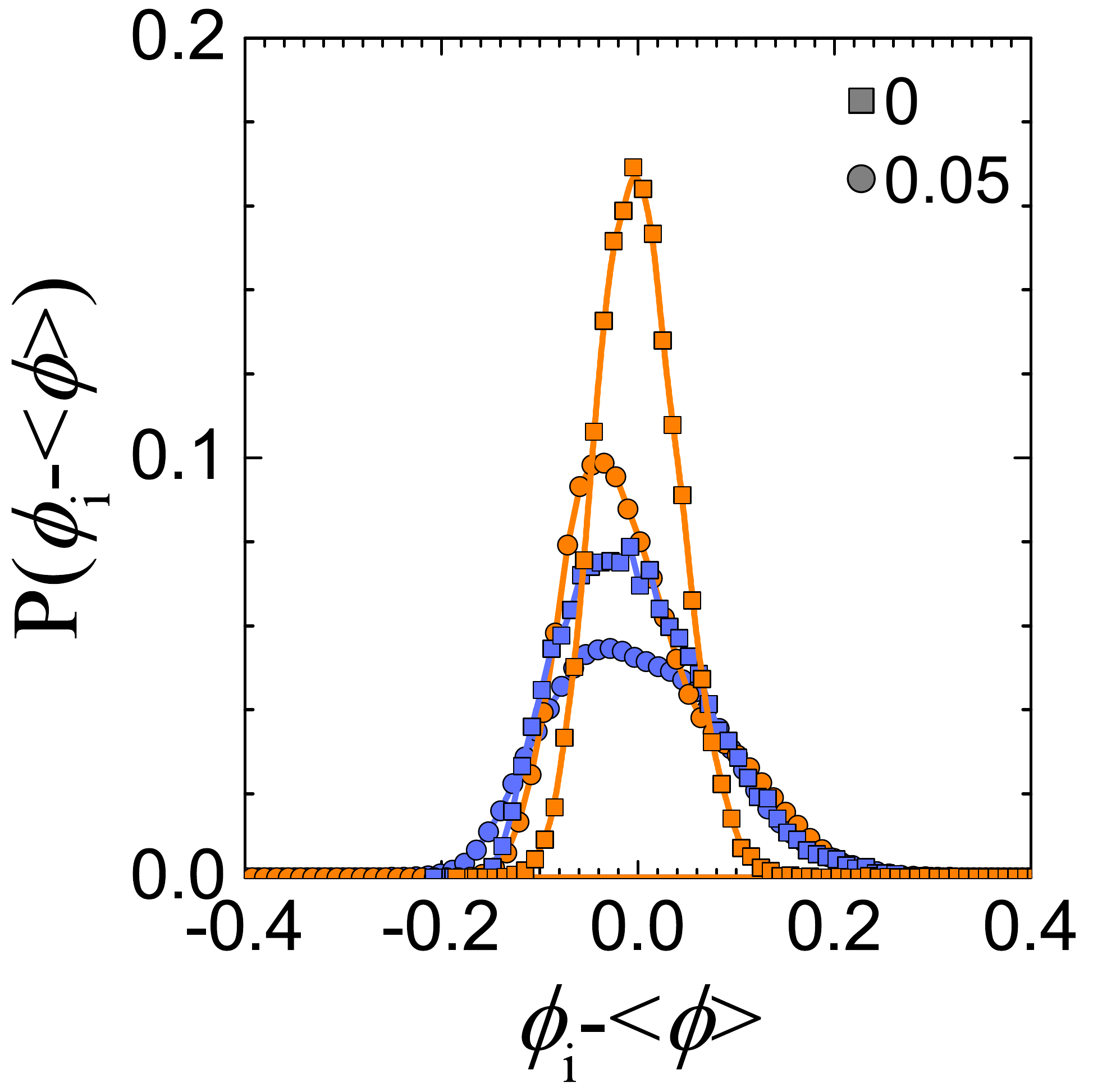}
\caption{Probability distributions of volume fraction $P(\phi_l)$ and $P(\phi_s)$ in experiments for two values of $c_p/c_p^\ast$. Orange data denote small particles $i=s$ and blue large $i=l$. Note that the methodology for the experiments leads to quantitatively different distributions from simulation.
}
\label{figIntensityFluctuationsExperiment}
\end{figure}

Figure~\ref{figIntensityFluctuationsSimulation} presents our simulation results on the critical isochore which show that $P(\phi_s)$ is essentially Gaussian, while for the state closest to the critical point ($c_p/c_p^\ast=0.056$) (blue triangles) $P(\phi_l)$ is non-Gaussian with a distinct tail extending to higher values of $\phi_l$, \emph{i.e.} towards the critical point. Similar behaviour is found in the experiments (Fig.~\ref{figIntensityFluctuationsExperiment}). These results suggest that the fluctuations in the two species are different in their sensitivity to deviations from criticality: the large particles with their stronger interparticle attractions respond first to approaching criticality; the small particles with their weaker attractions only do so much closer to the critical point than we approach here.

\section{Conclusions} 
\label{sectionConclusions}

Using particle-resolved studies and bespoke Monte Carlo simulation we have investigated the phase behavior of a simple ternary mixture of two colloidal and one polymer species. We have recast this ternary system as a binary colloid mixture, with effective interactions between the particles obtained by integrating out the polymer degrees of freedom. The current theoretical understanding of such mixtures is limited. Here we see that adding a second colloidal species introduces a remarkable level of complexity into a well-understood system. Although a superficial inspection of Fig.~\ref{figBinaryPhaseImages} suggests that colloid liquid-liquid demixing may occur, our simulations show that this is illusory. Rather, strong fractionation of the large particles occurs and there is only a vapor-liquid type separation. Thus our combined experimental and computational approach resolves the intriguing phase behaviour of this simple mixture.

We find the character of this vapor-liquid transition is much richer than in systems with one colloidal species due to multiple interaction ranges and strengths. At shallow quenches, the larger particles strongly prefer the liquid phase, while the small ones show only a weak preference --- a phenomenon which can give the appearance of liquid-liquid demixing. However, for deeper quenches the small particles migrate strongly to the liquid, reducing the concentration of large particles and leading to composition inversion \emph{i.e.} a maximum in the concentration of large particles in the liquid phase. For the deepest quenches, a gel forms. Our study also shows that while criticality is a collective phenomenon of the mixture, for slightly off-critical conditions, density fluctuations are dominated by the larger colloids while the smaller species behave somewhat as ``spectators''. In other words, criticality and phase separation are driven predominantly by the large particles. Given the basic nature of this system, we expect that this behavior may be found to apply widely in materials and formulations which are based on mixtures of colloids and polymers, such as cosmetics, foods and pesticides.

\subsection*{Acknowledgments}

This work was supported jointly by Bayer CropScience AG and the UK Engineering and Physical Sciences Research Council  through the award of an Industrial CASE award to IZ. CPR acknowledges the Royal Society for financial support, and EPSRC grant code EP/H022333/1 for provision of equipment used in this work, and the European Research Council under the FP7 / ERC Grant agreement n$^\circ$ 617266 and Kyoto University SPIRITS fund. RP acknowledges the Development and Promotion of Science and Technology Talents Project (DPST) of Thailand. NBW acknowledges EPSRC Grants No. EP/F047800 and No. EP/I036192.  RE acknowledges support from the Leverhulme Trust under EM-2016-031. This research made use of the Balena High Performance Computing Service at the University of Bath.

%\bibliography{cccp}


\end{document}